\documentclass{article}
\usepackage{spconf,amsmath,graphicx,amsfonts,amssymb,todonotes,algorithm,algpseudocode,booktabs,cite,pdfpages}
\usepackage{multirow}

\title{Balanced SNR-Aware Distillation for Guided Text-to-Audio Generation}

\name{Bingzhi Liu$^{1,2,3}$, Yin Cao$^{4}$, Haohe Liu$^{5}$, Yi Zhou$^{1,2,3}$}

\address{School of Communication and Information Engineering, Chongqing University of Posts and\\ Telecommunications, Chongqing (CQUPT), China$^{1}$\\Chongqing Key Laboratory of Signal and Information Processing, CQUPT, Chongqing, China$^{2}$\\Intelligent Speech and Audio Research Lab (ISARL), CQUPT, Chongqing, China$^{3}$\\
Department of Intelligent Science, Xi'an Jiaotong-Liverpool University, China$^{4}$\\
Centre for Vision, Speech and Signal Processing, University of Surrey$^{5}$\\}

\begin{document}
\ninept
\maketitle

\begin{abstract}
Diffusion models have demonstrated promising results in text-to-audio generation tasks. However, their practical usability is hindered by slow sampling speeds, limiting their applicability in high-throughput scenarios. To address this challenge, progressive distillation methods have been effective in producing more compact and efficient models. Nevertheless, these methods encounter issues with unbalanced weights at both high and low noise levels, potentially impacting the quality of generated samples. In this paper, we propose the adaptation of the progressive distillation method to text-to-audio generation tasks and introduce the Balanced SNR-Aware~(BSA) method, an enhanced loss-weighting mechanism for diffusion distillation. The BSA method employs a balanced approach to weight the loss for both high and low noise levels. We evaluate our proposed method on the AudioCaps dataset and report experimental results showing superior performance during the reverse diffusion process compared to previous distillation methods with the same number of sampling steps. Furthermore, the BSA method allows for a significant reduction in sampling steps from 200 to 25, with minimal performance degradation when compared to the original teacher models.

% Diffusion models have shown promising result in the realm of text-to-audio generation tasks. However, their inherent limitation lies in slow sampling speeds, making them impractical for high-throughput applications. To overcome this challenge, progressive distillation methods have proven effective in producing more compact and efficient models. However, the method suffers from the problem of unbalanced weights both at high and low noise levels, which may affect the quality of the generated samples. In this paper, we adapt the progressive distillation method to TTA generation tasks and introduce a Balanced SNR-Aware(BSA) method, which is an enhanced loss-weighting mechanism for diffusion distillation. The BSA method weights loss in a balanced manner for both high and low noise levels. We evaluated our proposed method on AudioCaps. Experimental results show that during reverse diffusion process, the proposed BSA method outperformes previous distillation methods with the same sampling steps. In addition, the BSA method can reduce the sampling steps from 200 to 25 with little performance degradation compared with original teacher models.

\end{abstract}

\section{Introduction}
\label{sec:intro}

Diffusion-based text-to-audio (TTA) generation techniques have emerged as a promising approach to AI content creation\cite{huang2023make,schneider2023mo,wang2023audit,yuan2023text,ruan2023mm} that utilizes a diffusion model to transform text into high-quality audio\cite{wang2023audit,yang2023diffsound,huang2023make,liu2023audioldm,ghosal2023text} that matches the content being created, which is valuable in domains such as movie production and gaming, where immersive audio experiences are critical. However its inference requires many sampling steps, hence is slow, which hinders its potential utilization. To address this challenge, Salimans et al.\cite{salimans2022progressive} introduced the concept of progressive distillation, aimed to increase the sampling rate. However, their method tended to exhibit a bias towards an excessive focus on low noise levels during distillation, resulting in models that underperformed when subjected to high noise levels. Subsequent attempts, such as Hang et al.'s\cite{hang2023efficient} adjustment to the training loss weight function, alleviated the overemphasis on low noise levels during training. However, this method overlooked the long-term exposure of models to high noise levels in later stages of progressive distillation, thereby leading to a bias in the model's predictive capabilities.

In response to these challenges, we introduce a Balanced SNR-Aware (BSA) method to weights loss functions in a SNR balanced manner. BSA method mitigates the overemphasis on small noise levels while preserving the model's effectiveness in predicting audio samples in high noise levels. We incorporate this loss-weighting strategy in the framework of progressive distillation. On this framework we redefine the progressive distillation algorithm to make it suitable for text-to-audio production tasks. Without loss of generality, we perform progressive distillation on Tango \cite{ghosal2023text}, which is a language-guided diffusion-based TTA model. Experimental results show that our proposed method outperformes previous methods with same sampling steps ranging from 100 to 25. In addition, we can reduce the sampling steps from 200 to 25 with little performance degradation compared with original teacher models.

% Codes are released \footnote{Code will be released after the acceptance of the paper.}.
In summary, the contributions of this paper are as follows:

\begin{itemize}
\setlength{\itemsep}{0pt}
\item We provide a novel representation of how the progressive distillation method can be adapted to Text-to-Audio (TTA) generation tasks. This application significantly reduces sampling steps of latent diffusion models, making them more practical for high-throughput scenarios.
\item We propose a Balanced Signal-to-Noise-Ratio-Aware (BSA) method, which is a loss-weighting strategy that allows the model to balance the signal-to-noise ratio during distillation, which can effectively improve performance of distilled models.
\end{itemize}

We organize this paper as follows. Section~\ref{sec:related-works} presents related works on diffusion models and knowledge distillation. Section~\ref{ssec:method} describes the proposed Balanced Signal-to-Noise Ratio Perceived Loss. Section~\ref{sec:experiment} presents our experimental settings and results. Section~\ref{sec:conclusion} summarizes our work.

\begin{figure*}[htb]  
\centering  
\includegraphics[width=0.95\textwidth]{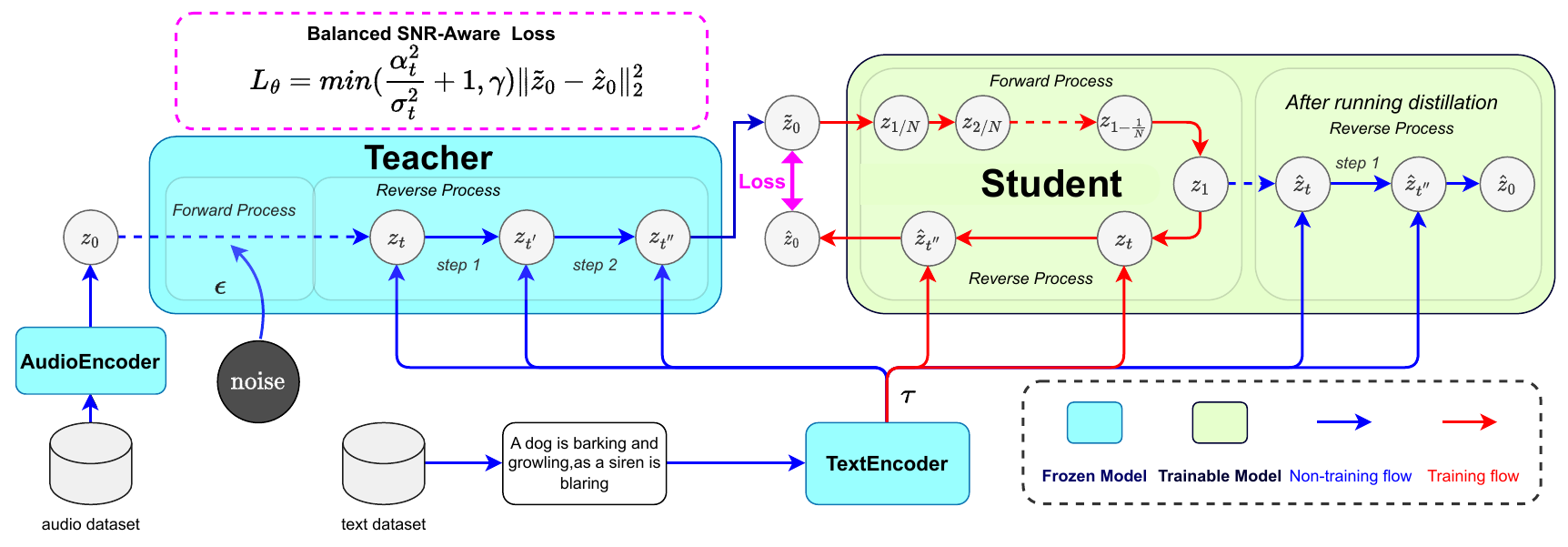}
\caption{In the process of distillation, we start with audio-text pairs from the dataset. After encoding the audio data and introducing noise in the forward process, a teacher model uses two steps of DDIM guided by the text data to predict $\tilde{z}_{0}$ from the noisy data $z_{t}$. This prediction $\tilde{z}_{0}$ becomes the target for the student model. The student model then generates $\hat{z}_{0}$ in a single step using text guidance, and this result is used to calculate the loss in conjunction with the teacher model's output. } 
\label{fig:kd} 
\end{figure*}

\section{Related Works}
\label{sec:related-works}

\subsection{Diffusion}
\label{ssec:diffusion}

For TTA generation, Diffsound\cite{yang2023diffsound} employed a discrete representation of the compressed Mel-spectrum to efficiently model the data, resulting in a generation speed five times faster than an autoregressive (AR) token decoder. However, the TTA generation process in Diffsound still involves hundreds of sampling steps. In contrast, AudioLDM\cite{liu2023audioldm} took a different approach, utilizing a latent-representation-based method. Their results showed that utilizing contrastive language audio pretraining (CLAP) as an audio encoder to allow audio-only data to be used in the training process can improve the computational efficiency of the TTA system. However, it is worth noting that the generation process of AudioLDM is still not improved. Tango\cite{ghosal2023text} utilized FLAN-T5\cite{chung2022scaling} to replace the previous CLAP. Although the quality of the generated samples has improved, there is still no improvement in generation speed.

\subsection{Knowledge Distillation}
\label{ssec:KD}

Hinton et al.\cite{hinton2015distilling} and Oord et al.\cite{oord2018parallel} introduced knowledge distillation, allowing for smaller and more efficient models while maintaining performance. Salimans et al.\cite{salimans2022progressive} proposed progressive distillation, reconstructing the loss weight function to train student models with one-step predictions matching the teacher model's two-step predictions. However, this approach lacks an upper limit on the loss weight function, leading to high sensitivity to minor changes in predictions, especially during initial training stages with high SNR. Hang et al.\cite{hang2023efficient} improved training by introducing an upper limit for the loss weight function, reducing the model's focus on low noise levels. However, they didn't establish a lower limit, causing challenges during later training iterations with high noise levels and low SNR. This increased uncertainty makes training less effective when faced with high noise levels in later stages.

In addition, in the field of text-to-audio generation, a distinctive characteristic sets it apart from image generation\cite{meng2023distillation,chang2023muse}: the utilization of an intermediary representation, where the choice of this representation, such as the  mel-spectrogra. It is important to underscore the absence of established progressive distillation methodologies tailored explicitly for text-to-audio generation.

\section{Method}
\label{sec:pagestyle}

\subsection{Latent diffusion model}
\label{ssec:method}

We employ diffusion models to approximate the genuine conditional data distribution $q(z_{0}\mid\tau)$, using a model distribution $p_{\theta} (z_{0}\mid\tau )$. Here, $z_{0}$ exists in a latent space, which is a compression of the mel-spectrogram $X \in \mathbb{R}^{T \times F}$ and is associated with an audio sample $x$. The dimensionality of $z_{0}$ is given by $C \times \frac{T}{r} \times \frac{F}{r}$, where $r$ signifies the compression level. Additionally, $C$ represents the number of channels in the latent representation, $X$ corresponds to the mel-spectrogram, and $T$ and $F$ denote the temporal and frequency dimensions, respectively. The text embedding $\tau$ is acquired through a pre-trained text encoder in FLAN-T5\cite{chung2022scaling}.

The diffusion model has two processes, a forward and a reverse processes. In the forward process\cite{song2020score}, the data distribution is systematically converted to a standard Gaussian distribution according to a predefined noise schedule\cite{liu2023audioldm}. In the reverse process, the audio $priori\ z_{0}$ is created through a stepwise denoising procedure with the guide text $\tau$. This denoising process is controlled by the following loss function\cite{ghosal2023text}:
\begin{gather}
L_{DM} = \sum_{n=1}^{N}\gamma_{n}\mathbb{E}_{\epsilon_{n}\sim \mathcal{N}(0,\boldsymbol{I}),z_{0}}\Vert \epsilon_{n}-\hat{\epsilon}_{\theta}^{(n)}(z_{n},\tau)\Vert^{2}_{2}
\end{gather}
where $N$ is the number of sampling steps. $z_{n}$ is obtained by adding noise $\epsilon_{n}\sim \mathcal{N}(0,\boldsymbol{I})$ to $z_{0}$. $\gamma_{n}$ is the weight of the loss function $n$\cite{hang2023efficient} and a measure of the SNR. This denoising process begins with an initial Gaussian distribution denoted as $p(z_{N})$, generating the audio $priori\ z_{0}$ step by step as follows\cite{ghosal2023text}:
\begin{gather}
p_{\theta}(z_{0:N}\mid\tau)=p(z_{N})\prod_{n=1}^{N}p_{\theta}(z_{n-1}\mid z_{n},\tau), \\
p_{\theta}(z_{n-1}\mid z_{n},\tau)= \mathcal{N}(\mu_{\theta}^{(n)}(z_{n},\tau),\Tilde{\beta}^{(n)}).
\end{gather}

The mean function and variance are parameterized as\cite{ghosal2023text}:
\begin{gather}
\mu_{\theta}^{(n)}(z_{n},\tau)=\frac{1}{\sqrt{\alpha_{n}}}\left[z_{n}-\frac{1-\alpha_{n}}{\sqrt{1-\Bar{\alpha}_{n}}}\hat{\epsilon}_{\theta}^{(n)}(z_{n},\tau)\right], \\
\Tilde{\beta^{(n)}}=\frac{1-\Bar{\alpha}_{n-1}}{1-\Bar{\alpha}_{n}}\beta_{n}.
\end{gather}

The structure of the diffusion model is adopted from U-Net\cite{ronneberger2015u}, which contains a cross-attention\cite{vaswani2017attention} module going to the input text guidance $\tau$.

\subsection{Progressive Distiallation for TTA}
\label{ssec:PD}

Figure 1 illustrates the asymptotic refinement process, where the only thing to be trained is the student model. The whole process differs from traditional training methods because the audio embedding $\tilde{z}_{0}$ used in the loss function during training is not the pure audio data $x$ from the dataset. Throughout the distillation process, the BSA method calculates the loss by comparing the results of the two sampling steps of the teacher model with the single sampling step of the student model, during which equilibrium in the model training is achieved by a loss weighting function, and then updating the student model. Iterating this process results in a model that produces similar output but with half the sampling steps of the teacher model.

Algorithm 1 illustrates the asymptotical refining process of progressive distillation for TTA in detail.A student model is initialized using the teacher model $\hat{x}_{\eta}(z_{t},\tau)$. let we start with $z_{t}$ till we reach $z_{t-1/N}$, where $N$ denotes the number of steps sampled by the student. We then perform the denoising procedure from $z_{t}$ to $z_{t-1/N}$, which allows the student model to learn a single step from teacher's two DDIM\cite{hang2023efficient} steps each time accompanying with a textual embedding $\tau$ input.  It is worth noting that given the teacher model and the starting point $z_{t}$ along with the textual embedding $\tau$, the prediction $\Tilde{x}(z_{t},\tau)$ from the teacher model is fully deterministic. However, with raw data $z_{0}$ and the textual embedding $\tau$ as inputs, the output of the forward process, $z_{t}$ is not fully deterministic due to multiple possible raw data inputs leading to the same noisy data $z_{t}$. This implies that the predictions of the original denoising model are a weighted average of $z_{0}$, leading to ambiguous predictions. Unlike the conventional way of training diffusion models, our approach provides both the teacher model and the starting point $z_{t}$. Consequently, the student model trained using our method can directly predict $z_{0}$, thereby avoiding intermediate steps, resulting in faster model sampling. As demonstrated in Section 4.4, our student model maintains the teacher model's performance while halving the number of steps. The student model then becomes the teacher model in the next distillation cycle.

\begin{algorithm}
\caption{Progressive distillation for TTA}\label{alg:cap}
\begin{algorithmic}
\Require Trained teacher model $\hat{x}_\eta(z_{t},\tau)$
\Require The matched text dataset $\mathcal{D}_{t}$ and audio dataset $\mathcal{D}_{a}$
\Require Loss weight function $w()$
\Require Student sampling steps $N$
    \For {$K$ iterations}
    \State $\theta \leftarrow \eta$ 
    \Comment{Init student from teacher}
    \While{not converged}
        \State $y \sim \mathcal{D}_{t},x \sim \mathcal{D}_{a}$
        \Comment{Textual description $y$, audio sample $x$}
        \State $\tau=f_{FLAN-T5}(y)$
        \Comment{$\tau$ is the output of FLAN-T5}
        \State $z_{0}=f_{AudioEncoder}(x)$
        \Comment{output of AudioEncoder}
        \State $t = i/N, i \sim Cat\left[1,2,\ldots,N\right]$
        \State $\epsilon \sim \mathcal{N}(0,\boldsymbol{I}) $
        \State $z_{t}=\alpha_{t}z_{0}+\sigma_{t} \epsilon$
        \Comment{Signal level $\alpha_{t}$, noise level $\sigma_{t}$ at time $t$}
        \State \# 2 steps of DDIM with teacher model
        \State $t^{'}=t-0.5/N, t^{''}=t-1/N$
        \State $z_{t^{'}}=\alpha_{t^{'}}\hat{x}_{\eta}(z_{t},\tau)+\frac{\sigma_{{t}^{'}}}{\sigma_{t}}(z_{t}-\alpha_{t}\hat{x}_{\eta}(z_{t},\tau))$
        \State $z_{t^{''}}=\alpha_{{t}^{''}}\hat{x}_{\eta}(z_{t^{'}},\tau)+\frac{\sigma_{{t}^{''}}}{\sigma_{t^{'}}}(z_{t^{'}}-\alpha_{t^{'}}\hat{x}_{\eta}(z_{t^{'}},\tau))$
        \State $\Tilde{z}_{0}=\frac{z_{t^{''}}-(\sigma_{t^{''}}/\sigma_{t})z_{t}}{\alpha_{t^{''}}-(\sigma_{t^{''}}/\sigma_{t})\alpha_{t}}$
        \Comment{Teacher $\tilde{z}_{0}$ target}
        \State $L_{\theta}=w(\gamma_{t})\mathbb{E}_{\epsilon,t,\Tilde{z}_{0}}\Vert \Tilde{z}_{0}-\hat{x}_{\theta}(z_{t},\tau)\Vert^{2}_{2}$\\
        \Comment{student prediction $\hat{x}_{\theta}(z_{t},\tau)$ is equivalent to $\hat{z}_{0}$}
        \State $\theta \leftarrow \theta-\gamma\bigtriangledown_{\theta}L_{\theta}$
    \EndWhile
    \State $\eta \leftarrow \theta$
    \Comment{Student becomes next teacher}
    \State $N \leftarrow N/2$
    \Comment{Halve number of sampling steps}
    \EndFor

\end{algorithmic}
\end{algorithm}

\subsection{Advantages of Balanced SNR-Aware Method}
\label{ssec:ABSM}

Ho et al.\cite{ho2020denoising} used Equation (6) as the loss function for the diffusion model:
\begin{gather}
L_{\theta} = \Vert \epsilon-\hat{\epsilon}_{\theta}(z_{t})\Vert^{2}_{2}=\frac{\alpha_{t}^{2}}{\sigma_{t}^{2}}\Vert x-\hat{x}_{\theta}(z_{t})\Vert^{2}_{2}.\label{XX}
\end{gather}
\cite{salimans2022progressive} showed that Equation (6) is unsuitable in distillation. During the distillation process, the signal-to-noise ratio steadily approaches zero.  In the extreme scenario where the signal level $\alpha_{t}$ approaches zero, the noise data $z_{t}$ ceases to convey any information about $z_{0}$, and the predicted value $\hat{\epsilon}_{\theta}(z_{t})$ no longer contains meaningful data-related information.

As the number of progressive distillation iterations increases, the span of each step widens, more noise is eliminated, and the information implicitly encoded in $z_{t}$ decreases. This information decrease adversely impacts the model's performance, necessitating the introduction of a reconstructive loss function. To address this challenge, \cite{salimans2022progressive} proposed two solutions:
\begin{gather}
L_{\theta} = max(\frac{\alpha_{t}^{2}}{\sigma_{t}^{2}},1)\Vert x-\hat{x}_{t}\Vert^{2}_{2},\label{XX} \\
L_{\theta} = (1+\frac{\alpha_{t}^{2}}{\sigma_{t}^{2}})\Vert x-\hat{x}_{t}\Vert^{2}_{2}.\label{XX}
\end{gather}

Both methods effectively handle the problem of loss function weights reaching zero. However, they exhibit a limitation: they result in large weights for the progressive distillation loss function during the early stages of high SNR levels. Consequently, the entire weight adjustment process becomes centered around the early training phases, leading to degraded model performance.

In response to this challenge, \cite{hang2023efficient} introduced a solution in Equation 9, which assigns maximum weights to the loss function to prevent the concentration of model training weights at high SNR. However, this approach does not address the problem of the loss function weights dropping to zero at low SNR. The importance of late-stage training is that the low information content in this phase leads to a denominator in the model prediction $\hat{x}_{\theta}(z_{t})=\frac{1}{\alpha_{t}}(z_{t}-\sigma_{t}\hat{\epsilon}_{\theta}(z_{t}))$ approaching zero, making the model highly sensitive to small perturbations.
\begin{gather}
L_{\theta} = min(\frac{\alpha_{t}^{2}}{\sigma_{t}^{2}},\gamma)\Vert x-\hat{x}_{t}\Vert^{2}_{2}.\label{XX} 
\end{gather}

To address these limitations, we propose our Balanced SNR-Aware(BSA) method as follows:

\begin{gather}
L_{\theta} = min(\frac{\alpha_{t}^{2}}{\sigma_{t}^{2}}+1,\gamma)\Vert x-\hat{x}_{t}\Vert^{2}_{2}.\label{XX} 
\end{gather}

In BSA method, a hyperparameter $\gamma$ is introduced, which is set to $5$ empirically. The proposed method weights the loss function in a balanced SNR manner. During training, when SNR is high, the weight locks to $\gamma$, which prevents excessive focus on low noise information. When SNR is low, the weight approaches 1, which prevents ignoring the high noise information.

\section{Experiments}
\label{sec:experiment}

\subsection{Datasets}
\label{ssec:datasets}
In this paper, we use the AudioCaps dataset\cite{kim2019audiocaps} for evaluating BSA. The dataset contains a training set with 49,286 audio-text pairs and a test set with 975 audio clips, each accompanied by five manually written subtitles, totaling 4,785 audio-text pairs. All the audio clips in the dataset are 10 seconds long and were collected from YouTube. These clips were edited to ensure a consistent 10-second duration and were paired with text by annotators.

\subsection{Experimental process}
\label{ssec:experimental process}
We assess the effectiveness of three different loss weight strategies: the conventional Loss Weight Strategy, the Min-SNR-$\gamma$ Loss Weight Strategy, and our proposed BSA, using the AudioCaps dataset. While we employ Tango\cite{ghosal2023text} during the experiments, it's important to note that our BSA is versatile and can be applied universally to latent diffusion models. We perform three iterations of the potential diffusion model, ensuring convergence at each iteration. Subsequently, we test the models generated by each strategy on the test set five times, with the aim of reducing experimental variability. Finally, to mitigate experimental errors, we construct a 95\% confidence interval around the test results. Detailed experimental results are presented in Section 4.4.

\subsection{Evaluation Metrics}
\label{ssec:evaluation metrics}
In this work, we use Frechet Audio Distance (FAD)\cite{kilgour2019frechet} and Frechet Distance (FD). FAD is the use of the Frechet Initial Distance (FID) metric for the audio domain, which is a perceptual metric capable of measuring the perceptual effects of a variety of distortions.FD denotes the similarity between the production sample and the similarity between the production and target samples.

\subsection{Results}
\label{ssec:results}

In Table 1, we conduct varying numbers of inference steps (200, 100, 50, and 25) using Tango's best model for generating data samples, as evaluated with objective metrics from the AudioCaps test set. We then use the BSA method to progressively distillate this model into student models with 100, 50, and 25 inference steps, respectively. For comparison, we use methods\cite{salimans2022progressive,hang2023efficient} to do the same work. Our student model consistently outperformed others on the AudioCaps dataset in terms of the objective metric FAD and FD for the same inference steps. Specifically, our 25-step student model generated samples with a FAD of 1.672, which is lower than the samples generated by models trained by other methods. This shows that the BSA method generates higher quality samples than the existing methods, and at the same time, the BSA method can achieve eight times faster sampling speed with a model before distillation.

Figure 2 illustrates the performance of various methods across distillation iterations. Notably, our approach consistently surpasses both the traditional DDIM sampler and the method introduced by Salimans et al. Following three distillation iterations, our model attains a FAD score of 1.672 for the 25-step model. In comparison, the DDIM sampler yields a FAD of 3.455, Salimans' method produces a FAD of 2.274 for the 25-step model, and Hang's method results in a FAD of 1.866. Our method consistently maintains robust and dependable performance during the evaluation.

\begin{table}\footnotesize
\centering
\label{table1}
\caption{Performance of the teacher model and the student model with different number of step, including the data from the 200-step teacher Our reported metrics encompass both FAD and FD, along with their respective 95\% confidence intervals.  The efficacy of the BAS method stands out at 25 steps. After just three iterations, it enables us to produce higher-quality samples than would have been possible otherwise, achieving FAD = 1.672 and FD = 21.847.} 
\begin{tabular}{c|c|cc}
\toprule[2pt]

\textbf{Model} & \textbf{Sampling steps} & \textbf{FAD}$\downarrow$  & \textbf{FD}$\downarrow$ \\ \midrule[1pt]
               & 200                 & 1.5706             & 21.3416            \\
Teacher\cite{ghosal2023text}        & 100                 & 1.694$\pm 0.015$             & 21.643$\pm 0.140$            \\
               & 50                  & 2.049$\pm 0.031$             & 22.523$\pm 0.268$            \\
               & 25                  & 3.455$\pm 0.041$             & 25.722$\pm 0.106$            \\ \midrule[1pt]
               & 100                 & 1.661$\pm 0.028$             & 21.627$\pm 0.188$            \\
Student\cite{salimans2022progressive}        & 50                  & 1.751$\pm 0.007$             & 21.981$\pm 0.324$            \\
               & 25                  & 2.274$\pm 0.174$             & 23.133$\pm 0.492$            \\ \midrule[1pt]
               & 100                 & 1.631$\pm 0.023$             & 21.794$\pm 0.303$            \\
Student\cite{hang2023efficient}        & 50                  & 1.711$\pm 0.014$             & 21.819$\pm 0.056$            \\
               & 25                  & 1.866$\pm 0.096$             & 22.147$\pm 0.492$            \\ \midrule[1pt]
               & 100                 & 1.593$\pm 0.026$             & 21.444$\pm 0.090$            \\
Student        & 50                  & 1.607$\pm 0.017$             & 21.706$\pm 0.156$            \\
(ours)      & 25                  & 1.672$\pm 0.052$             & 21.847$\pm 0.038$            \\ \bottomrule[2pt]
\end{tabular}
\end{table}

\begin{figure}[h]
\centering
\includegraphics[width=7.5cm,height=3cm]{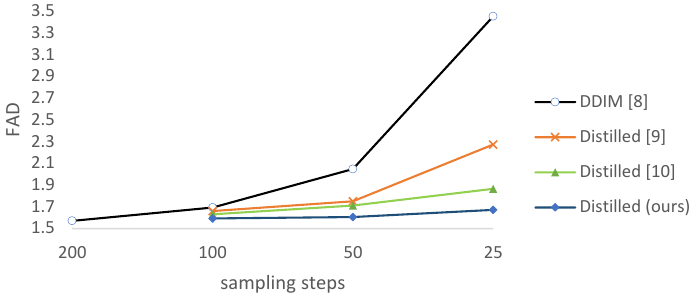}
\caption{Comparing the FAD of the various methods for training the model to convergence at the same number of sampling steps, the BSA method gives the least performance in exchange for the corresponding sampling rate at the 25-step result.}
\end{figure}

\section{Conclusions}
\label{sec:conclusion}

In this paper, we adapt progressive distillation to latent-diffusion-based Text-to-Audio (TTA) generation, marking it a practical implementation to distill knowledge from a TTA teacher model for a student model that can expedite sampling speed. We further propose a Balanced SNR-Aware (BSA) method that can balance the weights of a loss function so that the student model prevents excessive focus on low-noise information and ignorance of high-noise information. Through three iterations of distillation, we obtain a student model with only 25 sampling steps, which is eight times faster than the original teacher model. Experimental results show that the student model has little performance degradation compared to the teacher model and outperforms the previous best method by approximately 11.6\% in FAD and 1.3\% in FD.

\section{Acknowledgement}
\label{sec:print}
This work was supported in part by Grant ``XJTLU RDF-22-01-084''. For the purpose of open access, the authors have applied a Creative Commons Attribution (CC BY) licence to any Author Accepted Manuscript version arising.

\bibliographystyle{IEEEbib}
\bibliography{strings,refs}

\end{document}